\documentclass[fleqn,usenatbib]{mnras}
\usepackage[T1]{fontenc}
\DeclareRobustCommand{\VAN}[3]{#2}
\let\VANthebibliography\thebibliography
\def\thebibliography{\DeclareRobustCommand{\VAN}[3]{##3}\VANthebibliography}

\usepackage{amsmath}
\usepackage{amssymb}
\usepackage{mathtools}
\usepackage{comment}
\usepackage{hyperref}
\hypersetup{
    colorlinks = true,
    linkcolor = blue
    }
\usepackage[
        noabbrev,
        capitalise,
        nameinlink,]{cleveref}
\usepackage{multirow}
\usepackage[utf8]{inputenc}
\usepackage{float}
\usepackage{url}
\usepackage[table]{xcolor}
\usepackage{booktabs} 
\usepackage{multicol}        
\usepackage{bm}		
\usepackage{pdflscape}	
\usepackage{subfigure}
\usepackage{caption}
\usepackage{textcomp}
\usepackage{array}
\usepackage{nccmath}
 \usepackage{listings} 
\usepackage{alltt}
\usepackage{comment}
\usepackage{float}
\usepackage{romannum}
\graphicspath{ {./Figs/} }
\newcommand{\orcid}[1]{\href{https://orcid.org/#1}{\includegraphics[width=12pt]{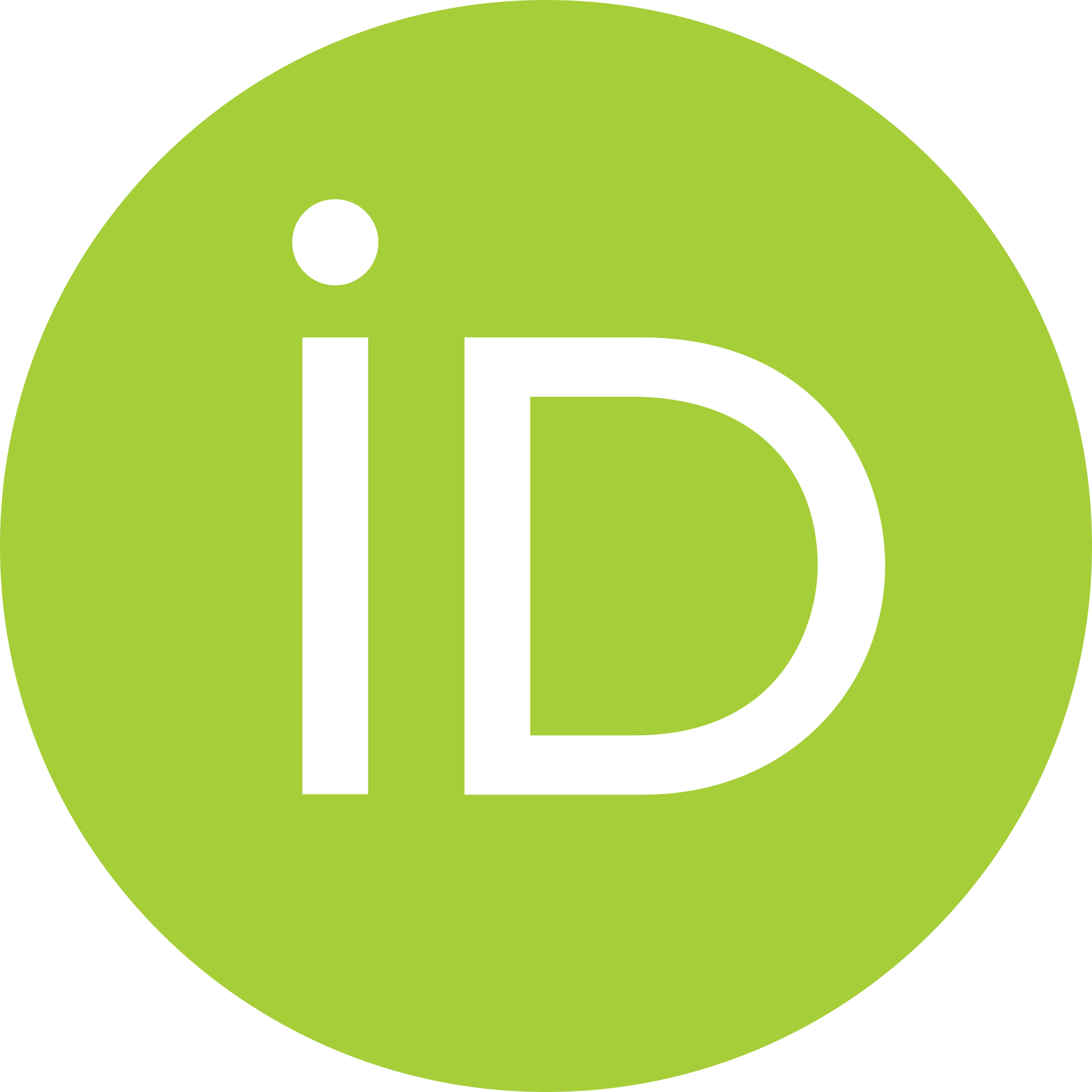}}}
\usepackage[normalem]{ulem}
\definecolor{navy}{RGB}{0, 0, 128}

\title[Probing clustering dark energy using deep learning]{Clusternets: A deep learning approach to probe clustering dark energy}
\author[Amirmohammad Chegeni et al.]{\noindent
Amirmohammad Chegeni \orcid{0000-0003-0251-9319} $^{1}$\thanks{chegeni@strw.leidenuniv.nl},
Farbod Hassani \orcid{0000-0003-2640-4460} $^{2}$\thanks{farbod.hassani@astro.uio.no},
Alireza Vafaei Sadr \orcid{0000-0002-5733-6678} $^{3,4,5}$\thanks{alireza.vafaeisadr@unige.ch},
\vspace{-4pt} 
Nima Khosravi \orcid{0000-0001-5723-2580} $^{1,6}$\thanks{n-khosravi@sbu.ac.ir}, \and
and Martin Kunz \orcid{0000-0002-3052-7394} $^{5}$ \thanks{martin.kunz@unige.ch}
\vspace{8pt}
\\ 
$^{1}$Department of Physics, Shahid Beheshti University, 1983969411, Tehran, Iran\\
$^{2}$Institute of Theoretical Astrophysics, Universitetet i Oslo, 0315 Oslo, Norway \\
$^{3}$Department of Public Health Sciences, College of Medicine, The Pennsylvania State University, Hershey, PA 17033, USA\\
$^{4}$Institute of Pathology, University Hospital RWTH Aachen, Pauwelsstra{\ss}e 30, 52074 Aachen, Germany\\
$^{5}$Department de Physique Theorique and Center for Astroparticle Physics, University of Geneva, 1211 Geneva, Switzerland\\
$^{6}$Department of Physics, Sharif University of Technology, Tehran 11155-9161, Iran\\
}

\date{Accepted XXX. Received YYY; in original form ZZZ}

\pubyear{2024}

\begin{document}
\label{firstpage}
\pagerange{\pageref{firstpage}--\pageref{lastpage}}
\maketitle
\pagenumbering{arabic}
\begin{abstract}
Machine Learning (ML) algorithms are becoming popular in cosmology for extracting valuable information from cosmological data. 
In this paper, we evaluate the performance of a Convolutional Neural Network (CNN) trained on matter density snapshots to distinguish clustering Dark Energy (DE) from the cosmological constant scenario and to detect the speed of sound ($c_s$) associated with clustering DE. We compare the CNN results with those from a Random Forest (RF) algorithm trained on power spectra.
Varying the dark energy equation of state parameter $w_{\rm{DE}}$ within the range of -0.7 to -0.99, while keeping $c_s^2 = 1$, we find that the CNN approach results in a significant improvement in accuracy over the RF algorithm. The improvement in classification accuracy can be as high as 40\% depending on the physical scales involved. 
We also investigate the ML algorithms' ability to detect the impact of the speed of sound by choosing $c_s^2$ from the set $\{1, 10^{-2}, 10^{-4}, 10^{-7}\}$ while maintaining a constant $w_{\rm DE}$ for three different cases: $w_{\rm DE} \in \{-0.7, -0.8, -0.9\}$.
Our results suggest that distinguishing between various values of $c_s^2$ and the case where $c_s^2=1$ is challenging, particularly at small scales and when $w_{\rm{DE}}\approx -1$. However, as we consider larger scales, the accuracy of $c_s^2$ detection improves. Notably, the CNN algorithm consistently outperforms the RF algorithm, leading to an approximate 20\% enhancement in $c_s^2$ detection accuracy in some cases.
\end{abstract}




\section{Introduction}\label{sec1}
The standard model of cosmology ($\Lambda$CDM) 
is in agreement with many
observations, including anisotropies in the Cosmic Microwave Background (CMB) \citep{Planck2018I,Planck2018V}, the Baryon Acoustic Oscillations (BAO) \citep{BOSS:2016}, and supernovae observations \citep{Supernova1998vns}. This model is based on the existence of the cosmological constant ($\Lambda$) and Cold Dark Matter (CDM). Despite its successes in explaining cosmological observations, the physical origins of these components are not explained within the $\Lambda$CDM model. In addition to the theoretical concerns, some observational anomalies have been pointed out in cosmological studies. For instance, the Hubble tension which indicates around 4$\sigma$ discrepancy between the late and early time measurements \citep{Divalentino:2021izs, Riess:2021jrx, Riess:2018uxu}. Some milder tensions/anomalies are also reported in the literature, for example, the growth rate ($\sigma_8$) tension \citep{DiValentino:2020vvd,Abdalla:2022yfr}, the BAO curiosities \citep{Aubourg:2014yra}, and the CMB anomalies \citep{Perivolaropoulos:2014lua} (see \citep{Perivolaropoulos:2021} for an overview). These tensions and anomalies may be a hint of beyond $\Lambda$CDM cosmology.

In recent years, cosmologists have proposed various alternatives to address the accelerating expansion of the universe \citep{Hu:2007nk,Amendola:2006we,Copeland:2006wr,Ishak:2018his,Banihashemi:2018has}, including clustering Dark Energy (DE), also known as $w-c_s^2-$CDM model. This model is characterized by the equation of state $w_{\rm DE}$ and the speed of sound $c_s$, and is a viable candidate for the late-time cosmic acceleration when $w_{\rm DE}$ is close to $-1$.
Clustering DE models can be thought as an effective field theory of $k$-essence \citep{Armendariz:2000nqq, Armendariz:2000ulo}, which involves a single scalar field that is minimally coupled to gravity. Clustering DE has been widely studied and implemented in some Boltzmann and $N$-body codes \citep{Lesgourgues:2011re, Lewis:1999bs, Zumalacarregui:2016pph, Hu:2014oga, Hassani:2019lmy, Adamek:2016zes, Hassani:2019wed, Hassani:2020buk}.

In the next few years, cosmological surveys such as Euclid \citep{EUCLID:2011zbd}, the Dark Energy Spectroscopic Instrument (DESI) \citep{DESI:2016fyo}, the Legacy Survey of Space and Time (LSST)  \citep{LSST:2012kar}, and the Square Kilometre Array (SKA) \citep{Santos:2015gra} will provide precise measurements of large-scale structures, including modes in the highly non-linear regime. The future data give us a great chance to test our cosmological models accurately, particularly those that aim to explain the late-time cosmic acceleration \citep{Tegmark:1997rp, Goobar:2011iv}. However, studying the non-linear regime analytically is a difficult task, which underscores the need for accurate numerical simulations \citep{Springel:2005mi, Bagla:1999tx, Adamek:2014xba, Teyssier:2001cp, ENZO:2013hhu, Hassani:2020rxd}. For cosmologists, a key challenge will be to determine the nature of the late-time accelerating expansion and whether the physical mechanism behind it is the cosmological constant or other theories of gravity \citep{Clifton:2011jh, Peebles:2002gy}.

Recently, Machine Learning (ML)-based methods have been introduced as important tools for cosmological applications \citep{Dvorkin:2022pwo, Navarro:2022twv, Kreisch:2021xzq, Ntampaka:2019udw}, demonstrating the potential to overcome some of the computational limitations of traditional techniques. These methods will be useful for identifying and categorizing cosmic sources \citep{khramtsov2019kids}, extracting information from images \citep{Navarro:2021rxp}, and optimizing observational strategies \citep{Fluri:2019qtp}. For instance, ML methods have been applied in the study of galaxy clustering \citep{Domingo:2022rvn}, supernovae \citep{ishida2019, Navarro:2021pkb}, strong and weak gravitational lensing \citep{Tewes:2018she, Cavaglia:2020qzp}, cosmological simulations \citep{Navarro:2022qqq, Perraudin:2019bxl}, gravitational waves \citep{Biswas:2013wfa, George:2017vlv, Morawski:2021kxv}, and the CMB maps \citep{Sadr:2020rje, Farsian:2020adf}.

In this direction, we utilise the Convolution Neural Network (CNN) \citep{a56} and Random Forest (RF) \citep{a54} algorithms to classify various clustering DE models based on their speed of sound $c_s$ and equation of state $w_{\rm DE}$, as compared to the cosmological constant scenario. 
 In this study, we use sub-boxes of matter density snapshots to train the CNN algorithm and matter power spectra extracted from these sub-boxes to train the RF algorithm. Subsequently, we use the confusion matrix to compare the accuracy of these two algorithms and to study the power of each algorithm in distinguishing clustering DE cosmologies parametrised by different clustering DE parameters. 
 Furthermore, we particularly explore the performance of the CNN algorithm in comparison to RF, especially on small scales, where non-linear structure formation is dominant.

 Previous works related to classifying different modified gravity models have employed different cosmological probes including; weak lensing maps \citep{Peel:2018aei, Ribli:2019wtw,Fluri:2018hoy}, and dark matter distribution \citep{Ravanbakhsh:2017bbi, Mathuriya:2018luj, Lazanu:2021tdl}. Also, some works used various approaches, from the basic statistics of matter distribution \citep{Baldi:2013iza, Peel:2018aly} to different ML algorithms \citep{Schmelzle:2017vwd, Arjona:2021mzf, Rivera:2019hqt, Mancarella:2020jyu}.
 
The paper is organised as follows. In \Cref{sec2} we discuss the simulations, the ML algorithms we employ and the evaluation criteria that we use to find the accuracy of the algorithms. Our results and their physical interpretations are discussed in \Cref{sec3}. Finally, we summarise and discuss the main results of this work in \Cref{sec4}.


\section{METHOD}\label{sec2}

In this section, we discuss the data, the ML algorithms, and the evaluation strategy that we use in this work. In \autoref{sim} we introduce the simulations we use to train the ML algorithms. We discuss the ML algorithms (CNN and RF) and the evaluation strategy that we use in this work in detail in \autoref{cnn}.

\subsection{Simulations} \label{sim}
In this paper, we utilize the $k$-evolution code \citep{Hassani:2019lmy} to investigate the clustering DE scenario. $k$-evolution is a relativistic $N$-body code based on \textit{gevolution} \citep{Adamek:2016zes, Adamek:2015eda}.
In $k$-evolution, the coupled equations for the $k$-essence scalar field, and dark matter particles are solved consistently. Furthermore, the $k$-essence is considered as an effective field that can be parameterised by the equation of state $w_{\rm DE}$ and the speed of sound $c_s$. The effective field theory of $k$-essence is also referred to as clustering DE which is a terminology we use in this paper. The $\Lambda$CDM simulations are based on \textit{gevolution} $N$-body code.

In \textit{gevolution} and $k$-evolution, matter and gravitational potentials are treated non-linearly and they are sourced by the dark energy density. In $k$-evolution, dark energy is implemented as an independent component and can form non-linear structures in response to non-linear matter clustering.
A lower speed of sound for clustering DE can lead to increased non-linear behaviour in the dark energy component \citep{Hassani:2019lmy}. However, when the equation of state of the dark energy is fixed and is close to -1, the effect of the speed of sound on matter power spectrum is relatively small, and it is hard to constrain clustering DE using changes in the matter power spectrum.

In our clustering DE simulations, we investigate a wide range of values for the equation of state and speed of sound. Specifically, we consider  $w \in \{-0.7, -0.8, -0.85, -0.9, -0.95, -0.97, -0.98, -0.99\}$ and $c_s^2 \in \{1, 10^{-2}, 10^{-4}, 10^{-7}\}$. Moreover, we consider the following fixed values for the cosmological parameters; the reduced Hubble parameter $h = 0.67556$, the spectral index $n_s = 0.9619$, the amplitude of scalar perturbations $A_s = 2.215 \times 10^{-9}$ at the pivot scale $k_p = 0.05  \, \rm{Mpc/h}$, the baryon and CDM component densities $w_{\rm b} = 0.02203$, $w_{ \rm CDM} = 0.12038$,
 and the CMB temperature $ T_{\rm CMB} = 2.7255 \, \rm{K}$. To examine the accuracy of the ML algorithms across different physical scales, we consider three different simulation box sizes: $L=128 \, \rm{Mpc/h}$, $L=512 \, \rm{Mpc/h}$, and $L=2048 \, \rm{Mpc/h}$.
 In our simulations, we use an equal number of lattice points and particles ($N_{\rm pcl} = N_{\rm grid}$). For the majority of our simulations, we fix the value of $N_{\rm grid}$ to $256^3$. However, to investigate the convergence of our results, we also consider a case where $N_{\rm grid} = 512^3$. Moreover, we conduct six simulations for each scenario, using different seed numbers. Taking into account the three box sizes, different $w_{\rm DE}$-$c_s^2$ parameters, and 6 unique seed numbers for each case, our study incorporates approximately 600 simulations in total.
\\
To compute the density field from the particle snapshots we utilise the \textit{Pylians3}\footnote{\url{https://pylians3.readthedocs.io/en/master/}} library. It is worth adding that the spatial resolution of the maps generated using \textit{Pylians3} is selected to match the grid resolution of the main particle snapshots.
The corresponding matter power spectra are calculated using the cloud-in-cell mass assignment method utilising the same library. We perform a consistency test to evaluate the validity of power spectra for the sub-boxes using the \textit{Pylians3}. We extract 2000 power spectra from sub-boxes of the main box using \textit{Pylians3}. Subsequently, we compute the mean and variance of the power spectra and compare them with the power spectrum derived from the primary density field snapshot. Our results indicate that the mean of the spectra coincides\footnote{It should be noted that using \textit{Pylians3} to extract power spectra from non-periodic sub-boxes can result in power leakage across different scales due to the convolution effect inherent in the use of fast Fourier transforms. Although this effect is not significant in our analysis when considering large enough sub-boxes, it becomes important when dealing with smaller sub-boxes. In that situation, it would be important to utilise a more suitable power spectrum estimator to mitigate any inconsistencies.} with the main spectrum within the wavenumber range covered by each sub-box. Moreover, as we consider smaller sub-boxes, the variation around the mean increases. Our choice of sub-box sizes in this study is motivated by ensuring that the variations around the mean remain manageable, thereby facilitating the machine learning process.

In our ML training process, we extract 2000 sub-boxes from each density snapshot, resulting in a total of 10,000 sub-boxes for each cosmology when all the seed numbers are taken into account. These sub-boxes may overlap and are subsequently fed as input for the algorithms during the training process. This particular number of sub-boxes has been determined based on the convergence of accuracy for the ML algorithms which is discussed in \Cref{RS}.
\\
We consider various cosmological scales by considering different simulation box sizes and sub-box cut fractions of $1/2$ and $1/4$.  To mitigate aliasing inaccuracies, we restrict the RF analysis to a portion of the power spectrum that is well below the Nyquist frequency. However, excluding higher wavenumbers results in a reduction of approximately 3\% in overall accuracy. Interestingly, it appears that the algorithms can still learn from these seemingly inaccurate wavenumbers, suggesting potential for improving accuracy with their inclusion.
\subsection{Machine Learning Algorithms}\label{cnn}
In our study, we initially considered several widely used ML algorithms for classification tasks. These include Logistic Regression \citep{a50}, Naive Bayes \citep{a51}, Decision Trees \citep{a52}, Support Vector Machine \citep{a53}, K-Nearest Neighbours \citep{Cunningham_2021}, and Random Forest \citep{a54}. We applied these algorithms to power spectra for some values in ($w-c_s^2$) parameter grid. Our results suggest that the RF has the best performance among these. That is why we chose the RF for the comparison with the CNN algorithm. It is worth noting that when considering the advantages of utilising ML techniques, it is important to compare them with more conventional methodologies. We introduce an approach for distinguishing clustering DE models with different equations of state from the $\Lambda$CDM scenario, termed as the simple classifier (SC). In this method, we compute the variance of the density field over the training sub-boxes and construct a variance distribution for each cosmology. Subsequently, we evaluate the density field variance of each test box, and measure the distance of individual sub-box variances from the mean of the variance distribution associated with each cosmology. If the variance of a sub-box closely aligns with the mean of the variance distribution for a specific model, we designate that cosmological model to the corresponding test sub-box. Following this, we establish a confusion matrix and assess the accuracy of the SC method against our machine learning techniques. This provides a baseline against which we can judge the performance of the ``black box'' machine-learning methods. The comparative analysis reveals that the SC method exhibits lower accuracy compared to the RF and CNN models, providing additional motivation for the study of ML methods in a cosmological context. For additional details, refer to \Cref{SC}.

\begin{itemize}

\item \textbf{Random Forests:} The RF \citep{Ball:2009wd, Carliles:2007nf, baron2019machine} is a widely used supervised algorithm for classification tasks. This algorithm has been used in numerous areas of science \citep{a52, a54}. In this work, we train the RF algorithm over the matter power spectra. The power spectra are computed from the matter density sub-boxes, which are also used to train the CNN algorithm. 
In order to optimise the performance of the RF algorithm, we adjust its hyperparameters. The RF algorithm relies on two primary hyperparameters: the number of trees and the maximum depth \citep{a54}. In our analyses, we construct a hyperparameter grid encompassing a range of values for the number of trees (from 10 to 2000) and the maximum depth (from 10 to 100). Subsequently, we evaluate the accuracy of the RF algorithm across this grid to determine the optimal hyperparameters. It is important to note that we adjust the hyperparameters for all cases discussed in \Cref{sec3}.

\vspace{0.25cm}
\item \textbf{Convolutional Neural Network:} CNN is a highly effective and widely used technique for object recognition and classification in the field of ML. These networks employ a hierarchical architecture consisting of multiple layers, including convolutional and pooling operations. Through this architecture, the CNN can automatically learn and extract meaningful features from input images, enabling them to process complex visual data with remarkable accuracy. The CNN has demonstrated exceptional performance in various computer vision tasks, including image recognition, classification, and generation. In our study, we use a CNN on three-dimensional (3D) simulation density maps to make the most of this technique in distinguishing clustering DE models with different speeds of sound and equations of state, compared to the cosmological constant scenario. The architectural design, parameter selection, and hyperparameter optimization of the CNN algorithm employed in our work are described in \Cref{appendix1}.
\end{itemize}

\subsection{Evaluation criteria}\label{acc}
To assess and compare the results of the two ML algorithms, it is crucial to establish suitable classification criteria. Since distinguishing clustering DE from the $\Lambda$CDM model can be considered a binary classification task, we employ the Confusion Matrix (CM) \citep{Ting2017} as an essential tool for analysing classification errors in our study. The CM provides insights into the performance of each ML algorithm utilised.
In the context of our work in \autoref{detection_w}, the CM allows us to determine the accuracy of classifying examples as clustering DE or $\Lambda$CDM models. Specifically, the CM consists of four categories: True Positive (TP), True Negative (TN), False Positive (FP), and False Negative (FN). A TP corresponds to an accurately classified clustering DE, while a TN signifies a correctly classified $\Lambda$CDM model. Conversely, an FP occurs when a $\Lambda$CDM model is mistakenly identified as clustering DE, and an FN arises when the ML algorithm fails to recognize clustering DE cases, incorrectly classifying them as $\Lambda$CDM models.
 The CM for our analysis is presented below, displaying the performance of the ML algorithms in distinguishing clustering DE from $\Lambda$CDM:

\vspace{0.5cm}

\begin{tabular}{cc|c|c|}
& \multicolumn{1}{c}{} & \multicolumn{2}{c}{Predicted Class} \\
& \multicolumn{1}{c}{} & \multicolumn{1}{c}{clustering DE} & \multicolumn{1}{c}{$\Lambda$CDM} \\
\cline{3-4}
\multirow{2}{*}[2ex]{\rotatebox[origin=c]{90}{Actual Class}} & clustering DE & TP & FN \\
\cline{3-4}
& $\Lambda$CDM & FP & TN \\
\cline{3-4}
\end{tabular}

\vspace{1.0cm}
However, achieving perfect accuracy in identifying every instance of clustering DE is impractical in real-world scenarios.
To evaluate the performance of an ML algorithm, a commonly used and straightforward statistic is ``accuracy'' which is computed using the CM. This statistic is computed by summing up the correct predictions for both the clustering DE and $\Lambda$CDM classes and dividing it by the total size of the test dataset,

\begin{equation}
    \rm{Accuracy} = \frac{\rm{TP} + \rm{TN}}{\rm{TP+FN + TN+ FP}}\cdot
    \label{accuracy_eq}
\end{equation}

\begin{figure*}
\includegraphics[width=\textwidth]{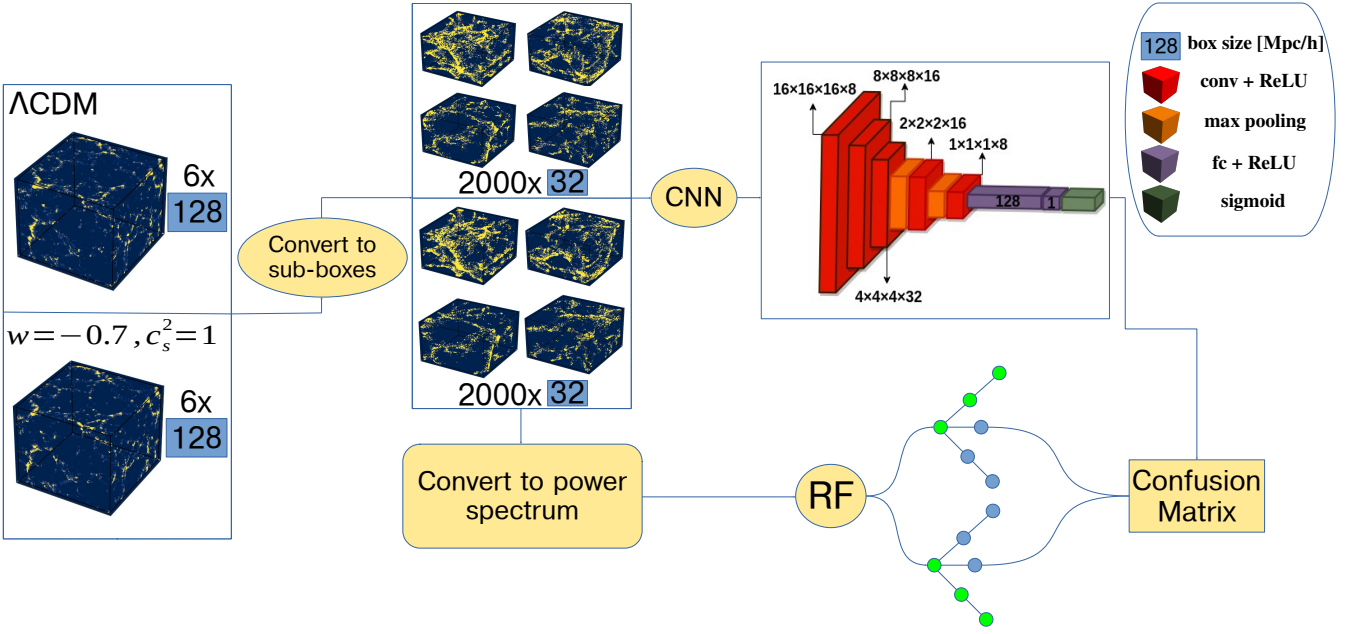}
\caption{We provide a schematic representation of our ML scheme for classifying clustering DE and the cosmological constant scenario, as discussed in \autoref{detection_w}. This scheme includes the CNN and RF algorithms trained on sub-boxes of both cosmological constant and clustering DE with $w_{\rm DE} = -0.98$ and $c_s^2=1$. The schematic illustrates the steps of extracting sub-boxes, calculating their power spectra as input for the RF algorithm, and assessing the accuracy of the algorithms using a confusion matrix.
\label{fig1}}
\end{figure*}

\vspace{0.0cm}
A schematic representation of our work is illustrated in \Cref{fig1}. In this figure, we show the details of the data preparation for each ML algorithm. In our study, we utilize six density snapshots with distinct seed numbers, representing different realizations of the same cosmology. To ensure robustness and generalization, we consider the sub-boxes extracted from five density snapshots as the training data, while the sub-boxes from the remaining density snapshot serve as the test and validation data. By incorporating results from five different seed numbers, we mitigate the potential influence of specific realizations and ensure the reliability of our results.
In our approach, we randomly select sub-boxes from each simulation and compute the matter power spectra for these sub-boxes. We then train a CNN algorithm using the 3D density maps from the sub-boxes and an RF algorithm using the corresponding matter power spectra. To assess the accuracy of each ML algorithm, we calculate the CM.
It is important to highlight that in the context of detecting the speed of sound ($c_s$), as discussed in \autoref{cs2_detection}, the classification task involves distinguishing between clustering DE with $c_s^2$ and clustering DE with $c_s^2=1$, while keeping $w_{\rm DE}$ fixed. 


\section{Results}\label{sec3}
In this section, we explore the accuracy of ML algorithms in distinguishing clustering dark energy theories by considering variations in the equation of state ($w_{\rm DE}$) and the speed of sound ($c_s$). We utilize the CNN algorithm for training on sub-boxes extracted from primary density snapshots, with varying simulation box sizes $L \in \{128, 256, 2048\}  \,\rm{ Mpc/h}$.
In all simulations, we utilize a fixed number of particles and grids, with $N_{\rm pcl} = N_{\rm grid} = 256^3$. Moreover, the snapshots are produced at a cosmological redshift of $z = 0$, which matches the current epoch in the Universe's evolution. 

In our study, we investigate sub-boxes that are either half or a quarter of the size of the main simulation box. These sub-boxes, along with the conditions $N_{\rm grid} = N_{\rm pcl} = 256^3$, play a crucial role in determining the length scales covered for each case analyzed.
For the case where the main box size is $L=2048 \,\rm{Mpc/h}$, and we cut sub-boxes that are half the size of the simulation box, the length scales span a range denoted as $\ell_{\frac12} \in [8, 1024] \,\rm{Mpc/h}$, which corresponds to a wavenumber range of $k_{\frac12} = {2 \pi}/{\ell_{\frac12}} \in [0.006, 0.680] \,\rm{h/Mpc}$.
Similarly, when we consider sub-boxes that are a quarter of the size of the main box, the length scales cover a range $\ell_{\frac14} \in [8, 512]$ Mpc/h, corresponding to a wavenumber range of $k_{\frac14} \in [0.012, 0.680]$ h/Mpc. In this context, $\ell$ represents the length scales incrementing by $\Delta \ell$, which is determined by the spatial resolution of the simulation $\Delta \ell = L/N_{\rm grid}$. The sub-index for $\ell$ indicates the portion of the main box that has been cut to form the sub-box. In \Cref{table0}, we present the length span and corresponding wavenumber ranges for each sub-box in the cases analyzed in our study. 

\begin{table*}
\begin{center}
      \begin{tabular}{ | c | c | c | c | c | }
    \hline
        \rowcolor{white}
   simulation box size &  \multicolumn{2}{c|}{length span} & \multicolumn{2}{c|}{ wavenumber range} \\ \hline
      & $\ell_{\frac12} [\,\rm{Mpc/h}] \in $ &  $\ell_{\frac14} [\,\rm{Mpc/h}] \in $ & $k_{\frac12} [\,\rm{h/Mpc}] \in $ &  $k_{\frac14} [\,\rm{h/Mpc}] \in $ \\ \hline
     $L = 2048 \,\rm{Mpc/h}$ &  [8, 1024]  &  [8, 512]  &  [0.006, 0.680]  & [0.012, 0.680]  \\  \hline
     $L = 512 \,\rm{Mpc/h}$ & [2, 256]  & [2, 128] &  [0.024, 2.720]  & [0.049, 2.720]  \\  \hline
     $L = 128 \,\rm{Mpc/h}$ & [0.5, 64]  & [0.5, 32] &  [0.098, 10.882]  & [0.196, 10.882]  \\  \hline
\end{tabular}
\end{center}
\caption{\label{table0} The length scale spans and corresponding wavenumber ranges for different simulation box sizes, as well as different cuts of the main simulation box (1/2 and 1/4 cuts). Throughout all cases, $N_{\rm grid} = 256^3$ remains constant, while we vary the box sizes and sub-box cuts to serve as input for training the ML algorithms. We determine the wavenumber ranges using the formulas: $k_{\rm min} = \frac{2 \pi}{\ell}$ and $k_{\rm max} = \sqrt{3} k_{{\rm N}} = \sqrt{3} \frac{\pi}{dx}$, where $k_{{\rm N}}$ is the Nyquist wavenumber and $dx$ is the spatial resolution.}
\end{table*}
Notably, the power spectra computed over sub-boxes of sizes $1/8$, $1/16$, and $1/32$ of the main boxes exhibit significant variation. Given the substantial variations observed in the power spectra of these sub-boxes, they proved unsuitable for the RF algorithm, as it would face challenges in effectively learning from such data. In contrast, we do not encounter this challenge when utilizing the CNN algorithm, which effectively learns from 3D density snapshots. The sub-box sizes of $1/2$ and $1/4$ of the main boxes used in this article provide the most consistent results when comparing the CNN and RF algorithms.

 The decision to use sub-boxes instead of main boxes is driven by the requirements of the RF and CNN algorithms. While this choice may lead to a reduced detectability of clustering DE due to the exclusion of certain large modes that are crucial for distinguishing the models, it aligns with the learning mechanisms of the algorithms. Both the RF and CNN algorithms rely on a large number of inputs to effectively learn the underlying patterns. Providing the full simulation box as input would necessitate training the algorithms with different simulations using different seed numbers to ensure convergence and learning from the patterns. However, such an approach would have significantly higher computational costs and make it prohibitively expensive to run these simulations.
 Therefore, utilizing sub-boxes for training the CNN and RF algorithms becomes a wise strategy. Cutting multiple sub-boxes from the main simulation box as a post-processing strategy offers a reasonable approach, striking a balance between excluding some modes and managing computational resources. It is important to note that by considering different sizes of main boxes, we effectively account for the modes that are lost by considering sub-boxes instead of full simulation boxes. 
 
\begin{figure}
\includegraphics[width=\columnwidth]{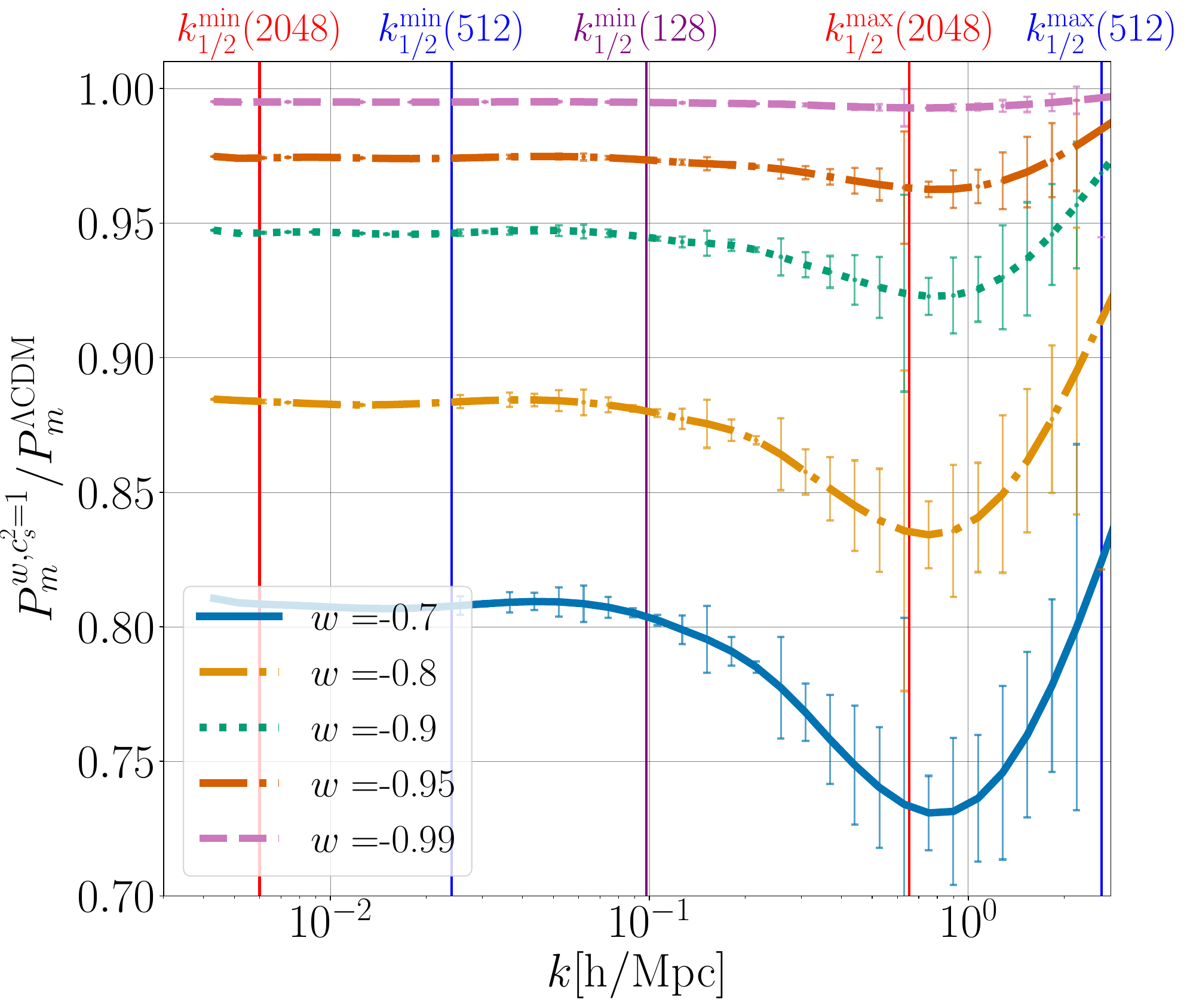}
\caption{ \label{fig4} The matter power spectra ratio of clustering dark energy scenarios for different equations of state, $w_{\rm DE}$, while keeping $c_s^2=1$, in comparison to the reference $\Lambda$CDM case. The figure includes vertical lines that indicate the range of wavenumbers covered ($k^{\rm min}_{1/2}(L)$, $k^{\rm max}_{1/2}(L)$) by sub-boxes obtained through a half-cut of the main simulation box, denoted by the $1/2$ sub-index. The sizes of the main simulation boxes ($L$) are indicated within parentheses, expressed in units of Mpc/h.  The error bars are computed by considering variations of $ P^{w, c_s^2=1}({\rm seed})/P^{\Lambda {\rm CDM} }({\rm seed}) $ computed for different seed numbers. }
\end{figure}

\begin{figure}
\includegraphics[width=\columnwidth]{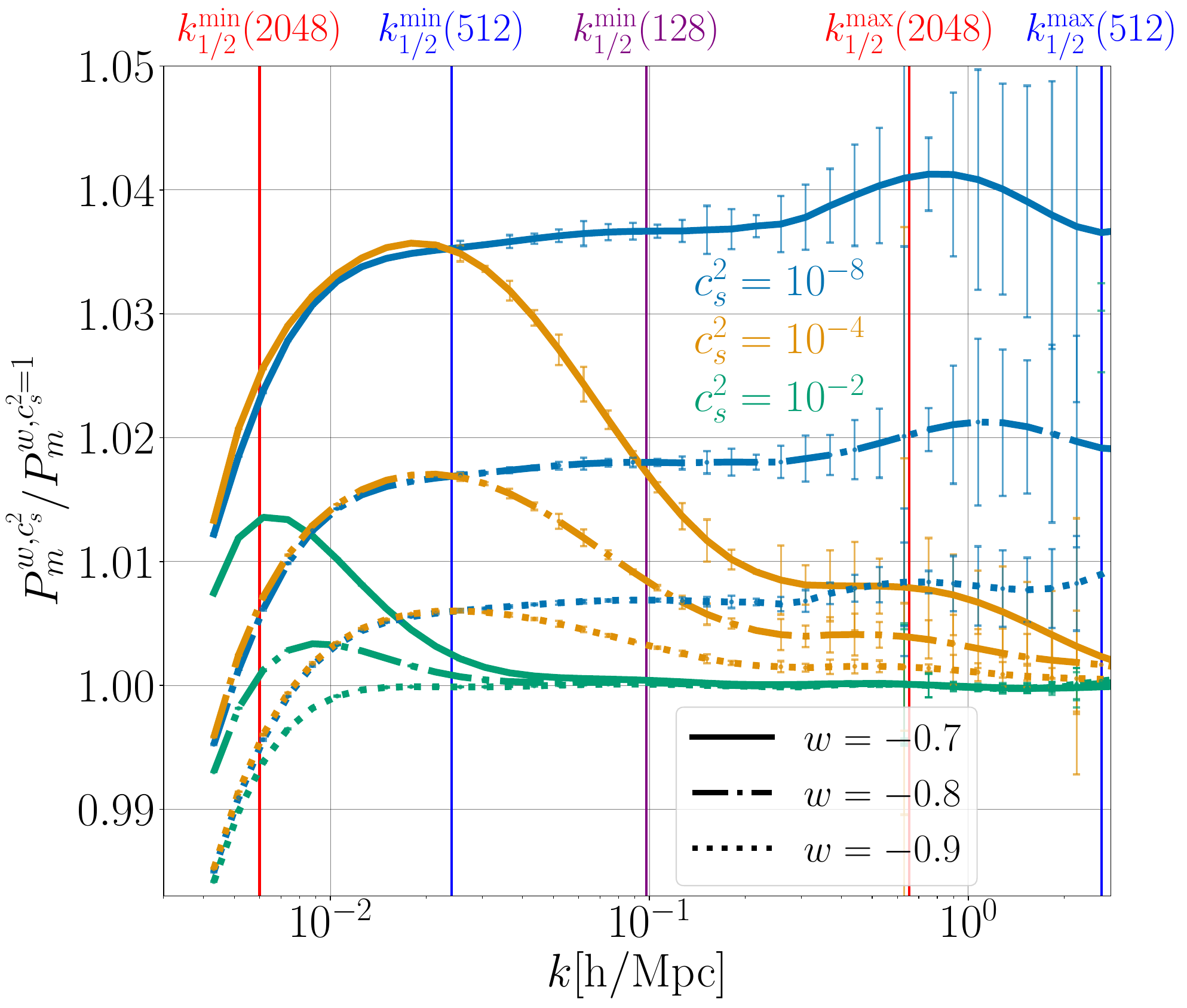}
\caption{ \label{fig5} The matter power spectra ratio, showing the comparison between different values of $c_s^2$ and $w_{\rm DE}$ with the case where $w_{\rm DE}$ remains constant and $c_s^2=1$, is illustrated in the figure. The vertical lines indicate the range of wavenumbers covered ($k^{\rm min}_{1/2}(L)$, $k^{\rm max}_{1/2}(L)$) by sub-boxes obtained through a half-cut of the main simulation box, denoted by the $1/2$ sub-index. The sizes of the main simulation boxes ($L$) are indicated within parentheses in units of Mpc/h. The error bars are computed by considering variations over the ratios of simulations with different seed numbers.} 
\end{figure}

In order to obtain initial insights into the behavior of different models before the training phase, we investigate the ratio of power spectra. \Cref{fig4} and \Cref{fig5} present the ratio of matter power spectra for the selected cases that are utilized for training the ML algorithms in the subsequent subsections. It is important to note that the RF algorithm is trained on the power spectra, not the ratios. The ratios presented in \Cref{fig4} and \Cref{fig5} are less noisy than the power spectra because cosmic variance is reduced in the ratio. Additionally, the ratios are obtained by averaging results from different seed numbers, which helps to reduce the fluctuations further, as reflected in the error bars. Finally, the parts of the simulations that are prone to resolution inaccuracies have been removed from the figures in the ratio. Therefore, the RF algorithm learns a much noisier power spectrum than what is shown in the ratios. This means that features that appear obvious in the ratios may be difficult to distinguish in the noisy power spectra.
 In \Cref{fig4}, we present the smoothed matter power spectra ratio of clustering dark energy with varying equation of state, $w_{\rm DE}$, and a fixed sound speed squared $c_s^2=1$, relative to the reference case of $\Lambda$CDM. On the other hand, \Cref{fig5} showcases the matter power spectra ratio of clustering dark energy with a fixed $w_{\rm DE}$ and different values of $c_s^2$, in comparison with clustering dark energy with the same $w_{\rm DE}$ but $c_s^2=1$.
 The figures include vertical lines indicating the wavenumbers covered by the sub-boxes, which are half the size of the main simulation box. These figures represent the average results of five simulations, each with different seed numbers. The error bars show how much the results vary around this average. Large error bars at small scales come from the use of a relatively small simulation box ($L=128$ Mpc/h). This limited box size leads to large variations in power spectra among different seed numbers and affects the ratios as well. 
The spectra exhibit significant variation in smaller box sizes across different seed numbers, because of the loss of long wavelength modes throughout the simulation due to cosmic variance.
  We combined the outcomes from three different box sizes, namely $L \in \{128, 256, 2048\} \,\rm{Mpc/h}$, while excluding the portion close to the Nyquist frequency. Furthermore, we applied the \textit{savgol} filter in \textit{Python} to achieve a smooth representation of the results.
 
These figures provide valuable insights into the scales (determined by box sizes and sub-box cuts) as well as the clustering dark energy parameters where the ML algorithms are expected to effectively distinguish different scenarios.  Analysing \Cref{fig4}, we observe that at large scales, the ratio remains relatively constant, with a suppression that is mainly determined by a factor related to $1+w_{\rm{DE}}$. However, at intermediate scales (around $k \sim 0.9$ h/Mpc), as non-linear effects become prominent, the ratio experiences a further decline. This decline arises from the enhanced growth of structures in the $\Lambda$CDM scenario compared to the $w$CDM scenario, leading to a more pronounced difference between the two cases.
At smaller scales, the ratio approaches unity (1) due to the dominance of matter non-linearities in both scenarios.

From \Cref{fig5}, it is evident that the ratio between different $c_s^2$ values and the case where $c_s^2=1$ increases as $1+ w_{\rm DE}$ increases. Furthermore, a smaller speed of sound leads to a more pronounced growth of matter structures. This is due to the fact that the sound horizon is located at larger wavenumbers or smaller scales. This effect leads to the amplification of clustering dark energy perturbations, resulting in a stronger clustering of matter within these specific regimes.

\subsection{Detecting Clustering Dark Energy:  $w_{\rm{DE}}$}\label{detection_w}
\begin{figure}
\includegraphics[width=\columnwidth]{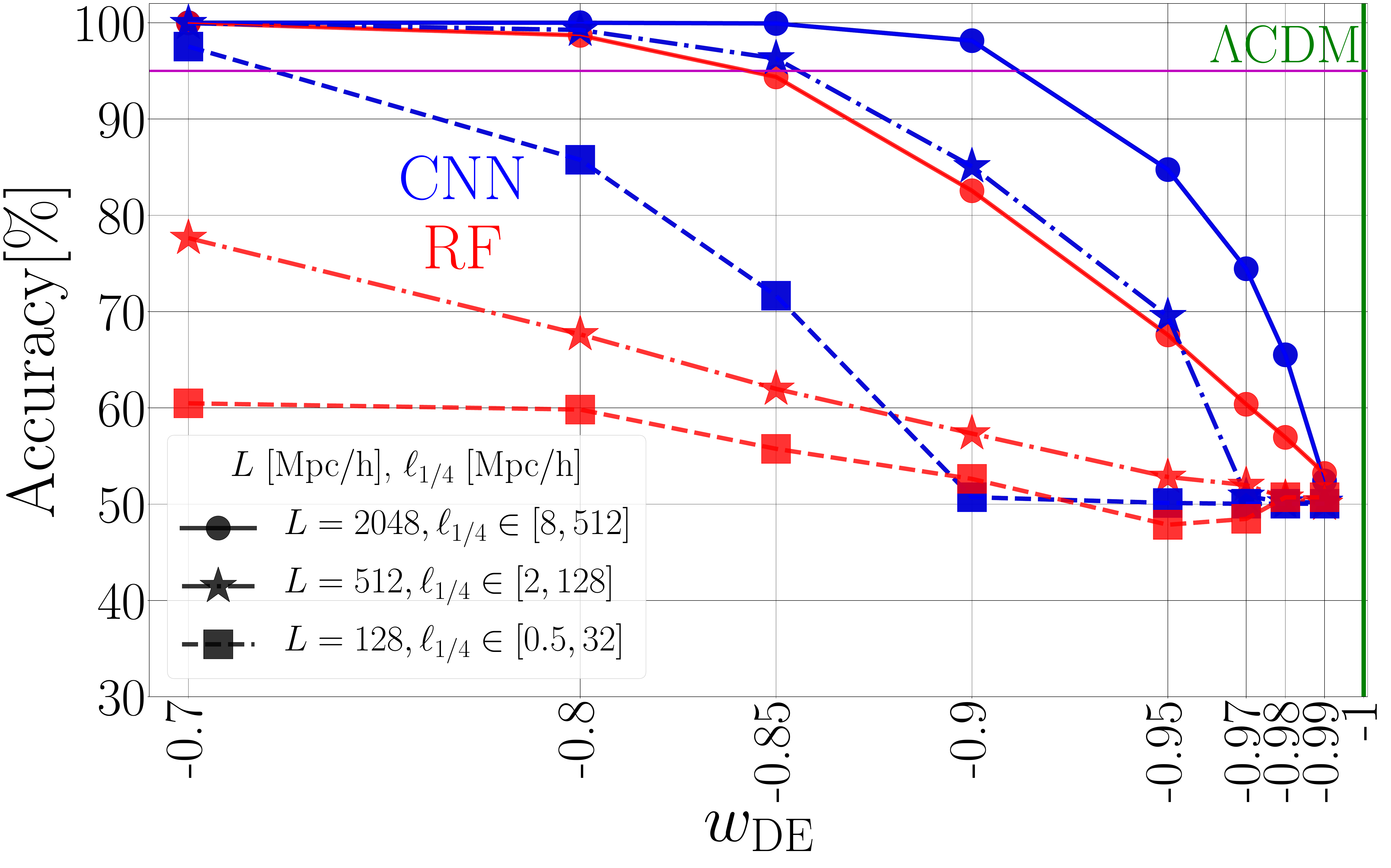}
\caption{ \label{fig2}
  The detection accuracy of the RF (in {\color{black}red}) and CNN (in {\color{black}blue}) algorithms for clustering DE across a range of equation of state parameters, $w_{\rm{DE}}$. The horizontal purple line represents the $95\%$ accuracy threshold. The data points on the plot correspond to different sizes of simulation boxes, allowing us to assess the performance of the RF and CNN algorithms across a range of scales. The analysis includes sub-boxes that are one-fourth ($1/4$) of the size of the main simulation box. The vertical green line on the plot represents the $\Lambda$CDM limit at $w_{\rm DE}=-1$. 
  \label{b2}}
\end{figure}

\begin{figure}
		\includegraphics[width=\columnwidth]{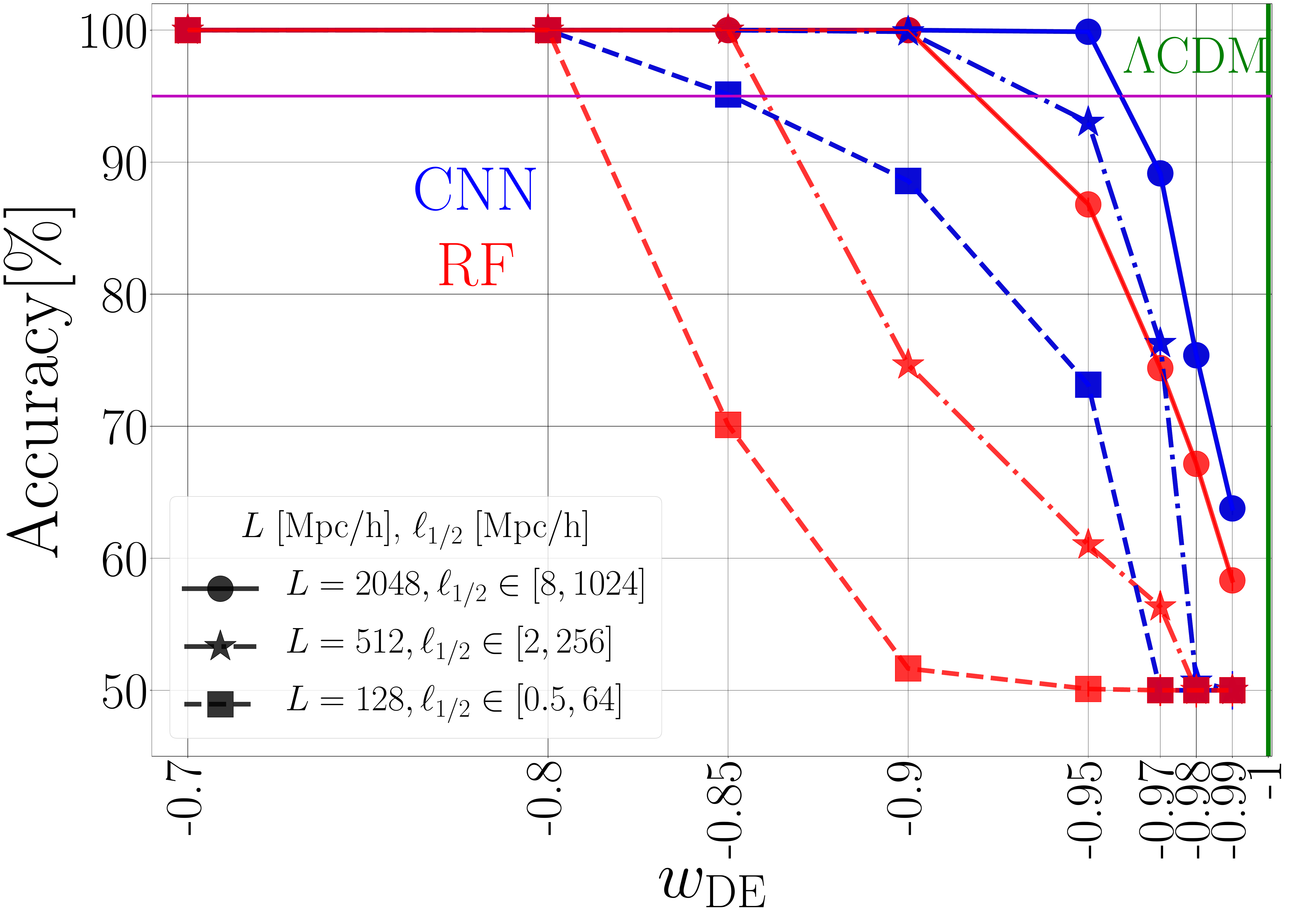}
		\caption{ \label{fig3} Same as \Cref{fig2} for the sub-boxes with 1/2 of the size of main simulation box.}
	\end{figure}

In this subsection, we explore the impact of various equations of state parameter ($ w_{\rm{DE}}$) for clustering DE, on the detectability of the CNN and RF algorithms. 
The ML algorithms are trained using simulation sub-boxes extracted from the main simulation box for different values of $w_{\rm{DE}}$ in  the set $\{-0.7, -0.8, -0.85, -0.9, -0.95, -0.97, -0.98, -0.99\}$, while $c_s^2 = 1$. As discussed previously, in the RF algorithm, the machine is trained using the power spectra of the extracted sub-boxes and in the CNN, we train it using 3D matter density snapshots. This training process is carried out for two different sub-box cut fractions, namely $1/2$ and $1/4$.
\\
In \Cref{fig2} and \Cref{fig3}, the accuracy of each algorithm in distinguishing clustering DE with different $w_{\rm{DE}}$ from the cosmological constant scenario ($w_{\rm{DE}}=-1$) is shown. The accuracies are plotted for different choices of simulation box size. In \Cref{fig2}, sub-boxes with a size of $1/4$ of the simulation box are utilized, while in \Cref{fig3}, sub-boxes that are $1/2$ of the simulation box are employed. In all the simulations, we maintain $c_s^2=1$, and $N_{\rm grid} = N_{\rm pcl} = 256^3$. To determine the length scale range and wavenumber ranges covered by each simulation box size and sub-box cut, refer to \Cref{table0}. In the figures, the blue lines represent the performance of the CNN algorithm, while the red lines show the accuracy of the RF algorithm for different values of $w_{\rm DE}$. The different point styles, such as stars, circles, and rectangles, correspond to different sizes of the simulation box, each corresponding to a distinct physical length span.
In \Cref{fig2} when considering the largest simulation box size of $L=2048 \,\rm{Mpc/h}$ which corresponds to $\ell_{\frac14} \in  [8, 512] \,\rm{Mpc/h}$ and $k_{\frac14} \in  [0.012, 0.680]  \,\rm{h/Mpc}$ (see \Cref{table0}), the CNN algorithm achieves an accuracy of $95\%$ in distinguishing clustering DE at approximately $w_{\rm{DE}} \approx -0.92$. On the other hand, the RF algorithm reaches this level of detectability at $w_{\rm{DE}}\approx -0.84$. 
 However, as we consider smaller simulation boxes, the accuracy of both ML methods tends to decrease. For a simulation box size of $L=512 \,\rm{Mpc/h}$ ($\ell_{\frac14} \in [2, 128] \,\rm{Mpc/h}$), the CNN algorithm achieves $95\%$ accuracy at around $w_{\rm{DE}}\approx-0.86$, and for $L=128 \,\rm{Mpc/h}$ ($\ell_{\frac14} \in [0.5, 32] \,\rm{Mpc/h}$), it achieves this level of detectability at approximately $w_{\rm{DE}}\approx -0.73$. However, it is important to highlight that the RF algorithm does not achieve the same level of accuracy for $w_{\rm{DE}} \le -0.7$, which is the range considered in our study.
 
By choosing larger sub-boxes as inputs for the ML algorithm analyses, specifically by cutting $1/2$ of the simulation box instead of using sub-boxes with a size of $1/4$, we can expect an improved accuracy in distinguishing clustering DE.
This improvement is attributed to the inclusion of a greater range of length scales in the analyses. However, employing larger sub-boxes also leads to larger data sets, which in turn increases the computational cost required for the algorithms to effectively learn from such extensive data.
 In \Cref{fig3}, we evaluate the accuracy of each ML algorithm using sub-boxes with a size equal to $1/2$ that of the main boxes. For the large simulation box $L=2048 \, \rm{Mpc/h}$ which corresponds to $\ell_{\frac12} \in [8, 1024] \,\rm{Mpc/h}$ and $k_{\frac12} \in  [0.006, 0.654] \,\rm{h/Mpc}$ (see \Cref{table0}), the CNN algorithm achieves $95\%$ accuracy at $w_{\rm{DE}}\approx-0.96$, while the RF algorithm reaches this level of accuracy at $w_{\rm{DE}}\approx-0.92$.
 Additionally, the CNN algorithm achieves $95\%$ accuracy  at $w_{\rm{DE}}\approx-0.94$ and $w_{\rm{DE}}\approx-0.85$ for $L=512\,\rm{Mpc/h}$ ($\ell_{\frac12} \in [2, 256] \,\rm{Mpc/h}$) and $L=128\,\rm{Mpc/h}$ ($\ell_{\frac12} \in [0.5, 64] \,\rm{Mpc/h}$), respectively, while the RF algorithm is only able to achieve this level of accuracy at $w_{\rm{DE}}\approx-0.86$ and $w_{\rm{DE}}\approx-0.81$. 

In summary, our results indicate that the CNN algorithm outperforms the RF algorithm in detecting the equation of state ($w_{\rm DE}$) of clustering DE across all scales. However, it is important to note that the accuracy of both methods decreases as the simulation box size decreases to the values considered in this study. 
The decrease in accuracy can be attributed to multiple factors, including the scale ranges present in the training set and how the effect of clustering DE manifests at different scales. Large variations among different seed numbers, as shown by the presence of large error bars at small scales in \Cref{fig4}, can complicate the learning process for algorithms trained on data from a small box size. 
 Furthermore, the complexity of structures also plays a role, becoming more evident at smaller scales due to non-linear matter clustering.
\\
It is worth noting that the difference in matter power spectra of clustering DE and $\Lambda$CDM is nearly constant\footnote{This constant change in amplitude may be degenerate with other parameters, such as $A_s$,  when in addition to the clustering DE, the cosmological parameters are varied.} at larger scales and decays at small scales ($k\ge0.9$ h/Mpc) as shown in \Cref{fig4}. This observation provides an explanation for the results we have obtained.

\subsection{Detecting Clustering Dark Energy: $c_s^2$} \label{cs2_detection}
The speed of sound, represented by $c_s$ is another key parameter for characterising clustering DE. It introduces a sound horizon for DE, where perturbations in DE decay on scales smaller than the sound horizon and grow on larger scales. So, small speeds of sound may result in non-linear clustering of DE. As discussed in \cite{Hassani:2019lmy}, the impact of the speed of sound on the matter power spectrum is negligible when $w_{\rm DE}$ is sufficiently close to $-1$.
The matter power spectrum's lack of sensitivity to the speed of sound poses an intriguing question about whether the CNN algorithm can enhance the detectability of clustering DE speed of sound by exploring different $c_s$ values while keeping $w_{\rm DE}$ fixed. Furthermore, the utilization of the RF algorithm on matter power spectra provides a more systematic analysis of the impact of the speed of sound on the matter power spectrum. By adopting these approaches, there is potential to enhance the constraints on $c_s^2$ in comparison to conventional methods.
\begin{table*}
\begin{center}
      \begin{tabular}{ | c | c | c | c | c | }
    \hline
        \multicolumn{5}{|c|}{$L = 512 \, \rm{Mpc/h}$, $w_{\rm{DE}}=-0.7$} \\ \hline
    length span &  \multicolumn{2}{c|}{$\ell_{\frac14} [\,\rm{Mpc/h}] \in  [2, 128]$} & \multicolumn{2}{c|}{$\ell_{\frac12} [\,\rm{Mpc/h}] \in  [2, 256]$} \\ \hline
         $c_s^2$ & CNN [\%] & RF [\%] & CNN [\%] & RF [\%] \\ \hline
     $10^{-2}$ & \cellcolor{pink} $50.98$ & \cellcolor{pink} $51.16$ & \cellcolor{pink} $50.9$ &\cellcolor{pink}  $52.15$  \\  \hline
     $10^{-4}$ & \cellcolor{pink} $50.65$ & \cellcolor{pink} $52.45$ & \cellcolor{pink} $54.6$ &  \cellcolor{pink}$56.47$  \\  \hline
     $10^{-8}$ &\cellcolor{yellow} $73.1$ & \cellcolor{pink} $55.4$ & \cellcolor{green} $95.75$ & \cellcolor{yellow}$61.32$  \\  \hline
\end{tabular}
     \vspace{-0.2cm}
      \begin{tabular}{ | c | c | c | c | c | }
    \hline
            \multicolumn{5}{|c|}{$L = 2048 \, \rm{Mpc/h}$, $w_{\rm{DE}}=-0.7$} \\ \hline
    length span &  \multicolumn{2}{c|}{$\ell_{\frac14} [\,\rm{Mpc/h}] \in  [8, 512]$} & \multicolumn{2}{c|}{$\ell_{\frac12} [\,\rm{Mpc/h}] \in  [8, 1024]$} \\ \hline
     $c_s^2$ & CNN [\%] & RF [\%] & CNN [\%] & RF [\%] \\ \hline
     $10^{-2}$ & \cellcolor{pink}$49.95 $ & \cellcolor{pink}$50.75 $ & \cellcolor{pink}$50.25 $ & \cellcolor{pink}$51.72$  \\  \hline
     $10^{-4}$ & \cellcolor{yellow}$68.35 $ & \cellcolor{yellow}$60.55 $ &\cellcolor{green} $94.87 $ & \cellcolor{yellow}$70.62$  \\  \hline
     $10^{-8}$ & \cellcolor{green}$91.7 $  & \cellcolor{yellow}$70.7 $ & \cellcolor{green}$100 $ & \cellcolor{green}$92.75$  \\  \hline
\end{tabular}
\end{center}
\caption{\label{table2} The accuracy of the CNN and RF algorithms is evaluated for different box sizes: $L=512 \,\rm{Mpc/h}$ on the left and $L=2048 \,\rm{Mpc/h}$ on the right, while keeping the equation of state fixed at $w_{\rm{DE}}=-0.7$, and varying the values of $c_s^2$. The accuracies are calculated by comparing the results with the reference case of $c_s^2=1$. Additionally, different sub-box sizes, representing 1/4 and 1/2 of the main boxes, are considered. The table also displays the corresponding length spans ($\ell_{\frac14}$ and $\ell_{\frac12}$) for each simulation cut and box size. The accuracy levels are visually represented using color coding: Pink cells indicate accuracy below 60\%, yellow cells indicate accuracy between 60\% and 80\%, and green cells indicate accuracy higher than 80\%.
}
\end{table*}
\begin{table*}
\begin{center}
      \begin{tabular}{ | c | c | c | c | c | }
    \hline
        \multicolumn{5}{|c|}{$L = 512 \, \rm{Mpc/h}$, $w_{\rm{DE}}=-0.8$} \\ \hline
    length span &  \multicolumn{2}{c|}{$\ell_{\frac14} [\,\rm{Mpc/h}] \in  [2, 128]$} & \multicolumn{2}{c|}{$\ell_{\frac12} [\,\rm{Mpc/h}] \in  [2, 256]$} \\ \hline
     $c_s^2$ & CNN [\%] & RF [\%] & CNN [\%] & RF [\%] \\ \hline
     $10^{-2}$ & \cellcolor{pink}$49.98 $ & \cellcolor{pink}$50.1$ & \cellcolor{pink}$50.98$ & \cellcolor{pink}$50.15 $  \\  \hline
     $10^{-4}$ & \cellcolor{pink}$50.04 $ & \cellcolor{pink}$51.65 $ & \cellcolor{pink}$50.02 $ & \cellcolor{pink}$50.45 $  \\  \hline
     $10^{-8}$ & \cellcolor{pink}$51.47 $ & \cellcolor{pink}$53.22 $ & \cellcolor{green}$81.57 $ & \cellcolor{pink}$51.30 $  \\  \hline
\end{tabular}
\vspace{-0.2cm}
      \begin{tabular}{ | c | c | c | c | c | }
    \hline
            \multicolumn{5}{|c|}{$L = 2048 \, \rm{Mpc/h}$, $w_{\rm{DE}}=-0.8$} \\ \hline
    length span &  \multicolumn{2}{c|}{$\ell_{\frac14} [\,\rm{Mpc/h}] \in  [8, 512]$} & \multicolumn{2}{c|}{$\ell_{\frac12} [\,\rm{Mpc/h}] \in  [8, 1024]$} \\ \hline
     $c_s^2$ & CNN [\%] & RF [\%] & CNN [\%] & RF [\%] \\ \hline
     $10^{-2}$ & \cellcolor{pink}$50 $ & \cellcolor{pink}$50.06 $ &\cellcolor{pink}$50.23$ & \cellcolor{pink}$50.76$  \\  \hline
     $10^{-4}$ & \cellcolor{pink}$58.68$ & \cellcolor{pink}$55.35 $ & \cellcolor{green}$81.72 $ & \cellcolor{yellow}$59.62$  \\  \hline
    $10^{-8}$ &\cellcolor{yellow}$76.29$  & \cellcolor{yellow}$59.93 $ &\cellcolor{green} $94.40 $ & \cellcolor{yellow}$74.25 $  \\  \hline
\end{tabular}
\end{center}
\caption{\label{table3} Same as \Cref{table2} for $w_{\rm DE}=-0.8$.}
\end{table*}
\begin{table*}
\begin{center}
\begin{tabular}{ | c | c | c | c | c | }
    \hline
        \multicolumn{5}{|c|}{$L = 512 \, \rm{Mpc/h}$, $w_{\rm{DE}}=-0.9$} \\ \hline
    length span &  \multicolumn{2}{c|}{$\ell_{\frac14} [\,\rm{Mpc/h}] \in  [2, 128]$} & \multicolumn{2}{c|}{$\ell_{\frac12} [\,\rm{Mpc/h}] \in  [2, 256]$} \\ \hline
         $c_s^2$ & CNN [\%] & RF [\%] & CNN [\%] & RF [\%] \\ \hline
     $10^{-2}$ & \cellcolor{pink}$50.02$ & \cellcolor{pink}$49.45$ & \cellcolor{pink}$50$ & \cellcolor{pink}$49.85$  \\  \hline
     $10^{-4}$ & \cellcolor{pink}$49.89$ & \cellcolor{pink}$50.35$ & \cellcolor{pink}$49.59$ & \cellcolor{pink}$50.15$  \\  \hline
     $10^{-8}$ & \cellcolor{pink}$50.64$ & \cellcolor{pink}$51.24$ & \cellcolor{pink}$50.14$ & \cellcolor{pink}$50.23$  \\  \hline
\end{tabular}
      \begin{tabular}{ | c | c | c | c | c | }
    \hline
            \multicolumn{5}{|c|}{$L = 2048 \, \rm{Mpc/h}$, $w_{\rm{DE}}=-0.9$} \\ \hline
    length span &  \multicolumn{2}{c|}{$\ell_{\frac14} [\,\rm{Mpc/h}] \in  [8, 512]$} & \multicolumn{2}{c|}{$\ell_{\frac12} [\,\rm{Mpc/h}] \in  [8, 1024]$} \\ \hline
     $c_s^2$ & CNN [\%] & RF [\%] & CNN [\%] & RF [\%] \\ \hline
     $10^{-2}$ & \cellcolor{pink}$50.04$ & \cellcolor{pink}$51.08$ & \cellcolor{pink}$50.83$ & \cellcolor{pink}$52.96$  \\  \hline
     $10^{-4}$ & \cellcolor{pink}$50.48$ & \cellcolor{pink}$52.15$ & \cellcolor{pink}$50$ & \cellcolor{pink}$53.6$  \\  \hline
     $10^{-8}$ & \cellcolor{pink}$52.19$  & \cellcolor{pink}$52.91$ & \cellcolor{yellow}$71.65$ & \cellcolor{yellow}$60.57$  \\  \hline
\end{tabular}
     \vspace{0.2cm}  
 \vspace{0.1cm}
\end{center}
\caption{\label{table4} Same as \Cref{table2} for $w_{\rm DE}=-0.9$.}
\end{table*}
We present the results of our investigation on distinguishing clustering DE with different speeds of sound values compared to the case where $c_s^2=1$ while keeping the equation of state $w_{\rm DE}$ constant. The results are summarised in \Cref{table2} (for $w_{\rm DE}=-0.7$), \Cref{table3} (for $w_{\rm DE}=-0.8$), and \Cref{table4} (for $w_{\rm DE}=-0.9$). 
Moreover, we consider $c_s^2 \in \{ {10^{-2}, 10^{-4}, 10^{-8}}\}$ and three different simulation box sizes: $L=2048 \, \rm{Mpc/h}$, $L=512 \, \rm{Mpc/h}$ and  $L=128 \, \rm{Mpc/h}$ along with sub-box sizes that are $1/4$ and $1/2$ of the main boxes. The corresponding length scales and wavenumbers for each sub-box choice are summarised in \Cref{table0}.

In the tables, we demonstrate the results for $L=2048 \, \rm{Mpc/h}$ and $L=512 \, \rm{Mpc/h}$ cases, as we did not observe any improvement in the accuracy of the CNN and RF algorithms for all the $c_s^2$ values when the box size $L = 128$ Mpc/h for both sub-box cuts ($1/2$ and $1/4$) considered. In other words, when considering scales of $\ell_{\frac12} \in [0.5, 64]$ Mpc/h or $\ell_{\frac14} \in [0.5, 32]$ Mpc/h, the ML algorithms were unable to distinguish between cases where $c_s^2 \neq 0$ and the case where $c_s^2=1$, even when $c_s^2$ was sufficiently small. The challenge in distinguishing the case with $c_s^2 = 10^{-8}$, where the power ratio in \Cref{fig5} exhibits a large difference at relevant scales ($k_{\frac 12} \in [0.098, 10.882]$ h/Mpc), can be attributed to the considerable variation in the data sets caused by the choice of a small box size. This variation is evident from the presence of large error bars at small scales in the figure. Moreover, the complexity of the underlying structures adds to the challenge of identifying patterns arising from the speed of sound.
\vspace{-0.25cm}
\paragraph*{$\bf{c_s^2 = 10^{-2}} \;\;\;$} In the case where $c_s^2 = 10^{-2}$, the sound horizon defined as $\ell_{s}= H_0/c_s$ is $\ell_{s} \approx 1878$ Mpc/h, or equivalently $k_s \approx 0.0033$ h/Mpc. Dark energy perturbations decay within scales $\ell \le \ell_{s}$, and the ratio of dark energy perturbations to matter perturbations drops significantly on scales below the sound horizon \citep{Hassani:2019lmy}. Consequently, even for the largest simulation box and largest sub-box cut, which corresponds to a length scale span of $\ell_{\frac12} \in [8, 1024] < \ell_{s}$, we expect a similar effect on matter structures as in the case where $c_s^2=1$. This makes it difficult for the RF and CNN algorithms to distinguish the effect of the speed of sound. Our results in \Cref{table2}, \Cref{table3}, and \Cref{table4} for both the CNN and RF algorithms, across all simulation box sizes and sub-box cuts, confirm that the ML algorithms cannot distinguish $c_s^2=10^{-2}$ from $c_s^2=1$. 
\vspace{-0.25cm}
\paragraph*{$\bf{c_s^2 = 10^{-4}} \;\;\;$} In this case, the sound horizon is $\ell_s  \approx 188$ Mpc/h and $k_s \approx 0.033$ h/Mpc. According to \Cref{table0}, in both cuts of box sizes with $L=512$ Mpc/h, the dark energy perturbations have almost decayed. As a result, it becomes challenging to distinguish this case from $c_s^2=1$ for this specific simulation box choice. According to the results presented in \Cref{table2}, \Cref{table3}, and \Cref{table4}, both the RF and CNN algorithms achieve an accuracy close to 50\% for $c_s^2=10^{-4}$ case and $L=512$ Mpc/h, with a slight improvement when $w=-0.7$ in \Cref{table2}.
However, when considering larger length scale spans, as shown in the tables for $L=2048$ Mpc/h, we observe an improvement in accuracies, particularly when employing a sub-box cut of $1/2$. This larger sub-box cut covers larger scales and captures more effects from dark energy perturbations.
As indicated in \Cref{table2}, the accuracy of the CNN and RF algorithms can reach 95\% and 71\%, respectively, for the case of $c_s^2=10^{-4}$, $L=2048$ Mpc/h, and a $1/2$ sub-box cut. The accuracy decreases significantly when using a $1/4$ sub-box cut. Moreover, the accuracy depends on the value of the equation of state $w_{\rm DE}$, and when $w_{\rm DE}$ is closer to $-1$, the accuracy decreases as shown in \Cref{table3}, where the accuracy of the CNN and RF decreases to 82\% and 60\%, respectively. While when $w_{\rm DE}=-0.9$ in \Cref{table4}, the accuracy reaches $\sim 50\%$, and $c_s^2=10^{-4}$ becomes undetectable by both ML algorithms.
\vspace{-0.25cm}
\paragraph*{$\boldsymbol{c_s^2 = 10^{-8}}$} When $c_s^2 = 10^{-8}$, the sound horizon is $\ell_s\approx 1.88$ Mpc/h and $k_s\approx 3.34$ h/Mpc. The RF and CNN algorithms exhibit improved accuracy compared to other sound speeds, even when $L=512 \, \rm{Mpc/h}$ and a $1/4$ cut is considered, as shown in \Cref{table2} and \Cref{table3}. However, when $w_{\rm DE}=-0.9$, the cases are indistinguishable, and the accuracy drops to approximately 50\% (\Cref{table4}). Similar to other speeds of sound, as $w_{\rm DE}$ approaches $-1$, the accuracy decreases, and using a sub-box cut of $1/2$ leads to increased accuracy. 
 Notably, when considering a $1/2$ cut for $w=-0.8$ in \Cref{table3}, the CNN achieves an enhanced accuracy of 82\%, while the RF fails to effectively distinguish the cases. This difference between the RF and CNN outcomes may be attributed to differences in their classification power and/or the information encoded in the data. It is generally expected that the density snapshots provide additional higher-order statistical information compared to the power spectra alone. However, in order to understand the exact causes of this significant difference, further investigations are necessary, particularly focusing on the impact of higher-order statistics in the data. 

In short, the detection of the speed of sound can be challenging, particularly as the equation of state $w_{\rm DE}$ approaches $-1$. Even with the CNN algorithm using 3D density snapshots, accurate detection remains difficult. However, our results indicate that employing larger sub-boxes and extending the length scale spans consistently enhances the accuracy of both ML algorithms. Specifically, when considering our largest simulation box size of $L=2048$ Mpc/h, both ML algorithms demonstrate improved performance in distinguishing cases with different $c_s^2$ values.  Furthermore, we observe that the accuracy further improves as the speed of sound $c
_s$ becomes smaller and deviates from $c_s=1$.

\section{Conclusion and Discussion} \label{sec4}

In this study, we examined how well machine learning algorithms, namely the Convolutional Neural Network (CNN) and Random Forest (RF), can detect clustering DE. We investigated different equation of state ($w_{\rm{DE}}$) and speed of sound ($c_s$) parameters using numerical $N$-body simulations, specifically the \textit{gevolution} and $k$-evolution codes, to train and test the algorithms.
\\
For the detection of clustering DE, we first focused on the accuracy of the ML algorithms in distinguishing different values of $w_{\rm{DE}}$ from the cosmological constant scenario ($w_{\rm{DE}} = -1$). We trained the CNN algorithm on 3D matter density snapshots extracted from sub-boxes of varying sizes, while the RF algorithm was trained on the power spectra computed from the sub-boxes. We evaluated the performance of the ML algorithms across a range of scales, from linear to non-linear. Our results showed that the CNN algorithm outperforms the RF algorithm in detecting clustering DE across almost all scales. For the largest simulation box size of $L=2048 \,\rm{Mpc/h}$, with a sub-box size of $1/4$ the main box, the CNN algorithm achieved an accuracy of $95\%$ in distinguishing clustering DE at $w_{\rm{DE}} \approx -0.96$, while the RF algorithm reached this level of detectability at $w_{\rm{DE}}\approx -0.84$. As the simulation box size decreased, the accuracy of both algorithms decreased.

We also investigated the impact of the speed of sound ($c_s$) on the detectability of clustering DE. Our results showed that distinguishing different values of $c_s^2$ from the case where $c_s^2=1$ was challenging for both the CNN and RF algorithms. This difficulty was particularly pronounced when considering small simulation sizes. The accuracy of the ML algorithms was highly dependent on the length scales considered. Larger sub-boxes, which covered a broader range of scales, generally led to improved accuracy in detecting the speed of sound. However, even with the use of 3D density snapshots in the CNN algorithm, accurate detection of $c_s^2$ remained difficult when $w_{\rm{DE}}$ approached $-1$. Interestingly, for the cases where $w=-0.7$ and $w=-0.8$, we achieved successful detection of the speed of sound effect for speeds $c_s^2 \le 10^{-4}$ when using a large simulation box size of $L= 2048$ Mpc/h, particularly with a sub-box cut of $1/2$. Moreover, we observed an approximate 20\% improvement in accuracy when employing the CNN compared to the RF algorithm.

In conclusion, our study demonstrated the potential of ML algorithms, particularly the CNN and RF algorithms, in detecting clustering DE. Furthermore, we studied the detectability of clustering DE through power spectra using the RF algorithm, which provides us a systematic study.
 The CNN algorithm consistently outperformed the RF algorithm across different scales and parameter ranges.  
 The improved accuracy of the CNN algorithm could be due to the inclusion of higher-order statistics in the density maps, in contrast to the limited two-point statistics offered by the power spectrum. Further investigations would be valuable in determining whether higher-order statistics in addition to two-point statistics is the primary factor behind the CNN algorithm's performance or if other factors play a role.
 
In this study, we focused on investigating the detectability of clustering DE by varying only the parameters $w_{\rm DE}$ and $c_s$, while keeping all other cosmological parameters fixed. It is important to note that the degeneracy between the clustering DE parameters and the cosmological parameters could be important, but it falls beyond the scope of our current paper. For a comprehensive analysis, it would be ideal to explore the entire parameter space, including parameters such as $A_s$, $n_s$, $\Omega_m$, $w_{\rm DE}$, $c_s$, and others, to evaluate how well the machine is capable of detecting clustering DE parameters in the presence of variations in all parameters. However, conducting such investigations would be computationally expensive.

To effectively use ML algorithms with data from cosmological surveys, such as galaxy catalogues, it is crucial to train them using realistic datasets like mock catalogues generated from snapshots of $N$-body simulations. This provides a better representation of actual observations compared to training only on matter density snapshots. By training the ML models on these simulated datasets, we can study how well they detect the parameters $w_{\rm DE}$ and $c_s$ in real observational data. This methodology gives us valuable insights into the possibility of detecting clustering DE in future cosmological surveys and opens doors for future advancements in the field.

\section*{Acknowledgements}
We would like to thank the anonymous referee for insightful comments, particularly for suggesting the inclusion of the simple classifier test. AC also would like to thank Doogesh Kodi Ramanah, Alessandro Renzi, and Shima Amirentezari for useful conversations and comments. FH thanks Julian Adamek for helpful comments about manuscript and useful discussions.  
This work is supported by the University of Oslo  computational facilities and a grant from the Swiss National Supercomputing Centre (CSCS) under project ID s1051. FH is supported by the Overseas Research Fellowship from the Research Council of Norway and the UNINETT Sigma2 $-$ the National Infrastructure for
High Performance Computing and Data Storage in Norway. MK acknowledges funding by the Swiss NSF.

\section*{DATA AVAILABILITY}
The codes utilised in this study can be accessed publicly on GitHub at the following repository:  \href{https://github.com/AM-Chegeni/Clusternets}{https://github.com/AM-Chegeni/Clusternets}. The simulation data that support the findings of this study are available upon request
from the authors.

\bibliographystyle{mnras}
\bibliography{biblio} 


\appendix

\section{Convergence Test}\label{RS}

In this section, we verify the stability of our ML algorithms. It is crucial to ensure that our models maintain a reliable level of accuracy as we increase the amount of data used for training the models. 
To achieve this, we utilise larger simulations, providing the ML algorithms with a larger amount of data. Our objective is to ensure that the accuracy of the ML algorithms remains stable, unaffected by issues such as over-learning or premature termination of learning (under-learning). Over-learning occurs when the model becomes overly complex and the machine only memorises the training data, resulting in high accuracy for the training data but poor performance on test data. Conversely, under-learning arises when the model is too simplistic and fails to capture the underlying patterns in the data, leading to low accuracy on both the training and test data. By evaluating our ML algorithms on a fixed sub-box size and physical resolution, we can ensure that the models neither over-learn nor under-learn, and maintain a consistent level of accuracy.

We specifically consider three simulations: two with a size of $L = 256$ Mpc/h and another with $L = 512$ Mpc/h. In two of the simulations, we use a spatial resolution of $1$ Mpc/h and employ grid and particle numbers of $N_{\rm grid} = N_{\rm pcl} = 256^3$ and $N_{\rm grid} = N_{\rm pcl} = 512^3$, respectively. In the third simulation, we use $L = 256$ Mpc/h and $N_{\rm grid} = N_{\rm pcl} = 512^3$. However, for the ML algorithms' input, we degrade the snapshots to $N_{\rm grid} = 256^3$ to match the spatial resolution of the other two simulations.
The purpose of degradation is to observe the effects of finite resolution inaccuracies on the performance of the ML algorithms. By averaging the high-resolution simulation, we obtain better and more reliable results at small scales. This approach allows us to assess how inaccuracies stemming from finite resolution, which are reduced in the degraded simulation, impact the performance of the ML algorithms.
To ensure fair comparisons, sub-boxes of the same physical size, $64$ Mpc/h, were extracted from all simulations and utilized as inputs for the ML algorithms. These sub-boxes correspond to $1/4$ and $1/8$ cuts of the main simulation box, respectively.
 The analysis presented in \autoref{b4} shows that the overall size of the simulation does not affect the accuracy of both the CNN and RF algorithms. Instead, their performance depends on the length scales encompassed by the sub-boxes (here $\ell \in [1, 64]$ Mpc/h). Interestingly, utilizing a degraded snapshot slightly improves the results, indicating more accurate treatment at resolutions closer to the simulation's resolution.
\\
These findings demonstrate that both algorithms consistently maintain a stable and reliable level of accuracy. The inaccuracies arising from finite resolution do not significantly alter the accuracies of the ML algorithms. Moreover, the results suggest that the ML algorithms employed in our simulations, conducted with a grid and particle number of $N_{\rm grid } = N_{\rm pcl} = 256^3$, are robust and not susceptible to issues of over-learning or under-learning.

\begin{table*}
\begin{center}
    \begin{tabular}{ | c | c | c | c | }
    \hline
      Number of grids (${N}_{\rm grid}$) & Box size (Mpc/h) & CNN accuracy & RF accuracy \\ \hline
       $256^3$  & $256$ & $96.35 $ & $62.3  $ \\ \hline
       $512^3$  & $512$ & $97.38 $ & $60.65  $ \\ \hline
      $512^3 \xrightarrow{\text{degraded}} 256^3$  & $256$ & $97.68 $ & $66.2  $ \\ \hline
\end{tabular}
  \caption{\label{b4} The accuracy of different ML algorithms for a fixed spatial resolution and sub-box size of $64 $ Mpc/h where the length scale span $\ell \in [1, 64]$ Mpc/h is fixed. We consider the clustering DE with $ w_{\rm{DE}} = - 0.8, c_s^2=1$ and the $\Lambda$CDM scenarios. In the last row, a simulation was performed with an initial grid size of $N_{\rm{grid}} = 512^3$ and subsequently degraded to $N_{\rm{grid}} = 256^3$.}
\end{center}
\end{table*}

\begin{figure}
\includegraphics[width=\columnwidth]{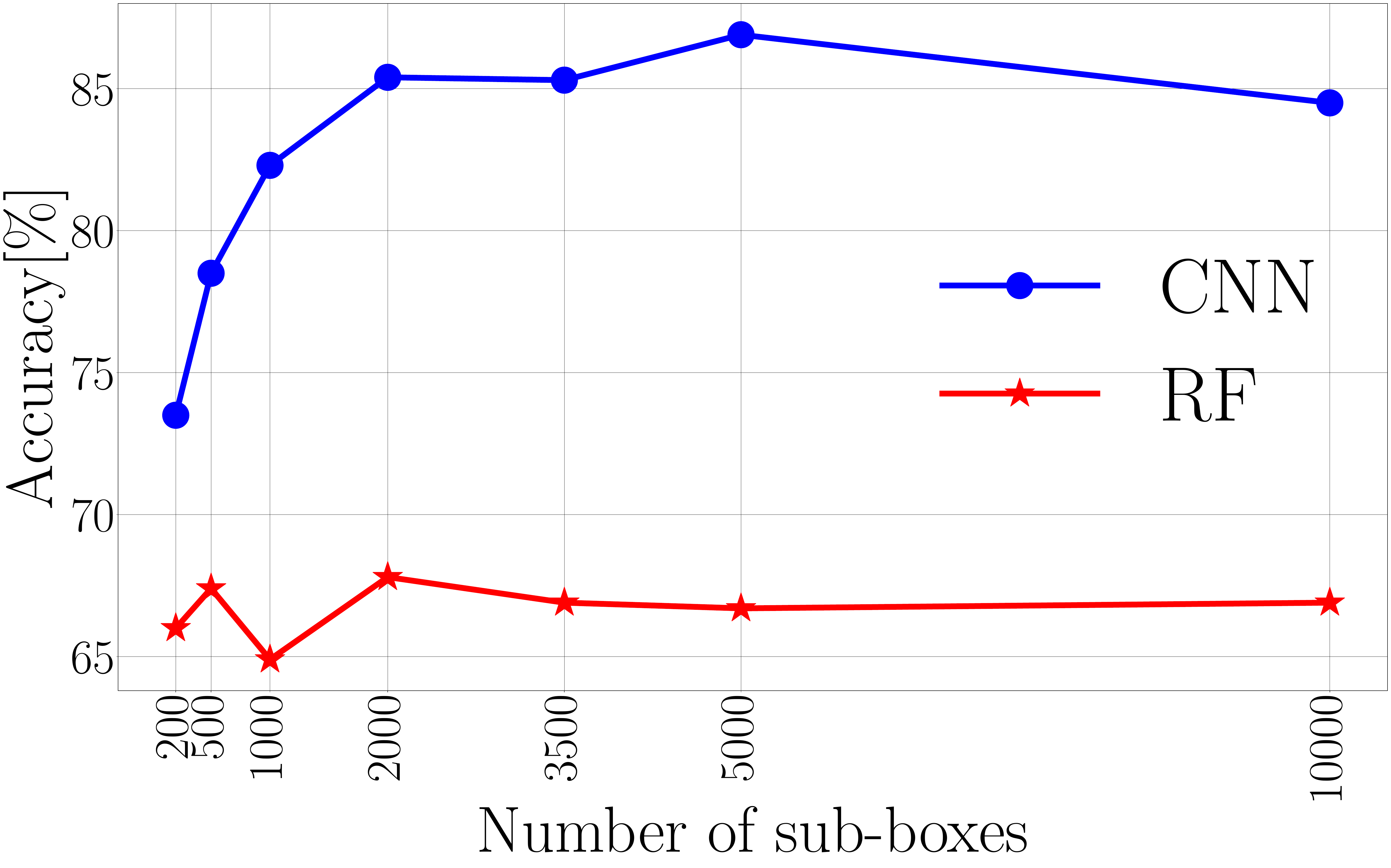}
\caption{ \label{fig6}
  The detection accuracy of the  CNN (in {\color{black}blue}) and RF (in {\color{black}red})  algorithms for different numbers of sub-boxes, from 200 to 10,000. 
We consider sub-boxes that have a size one-fourth (1/4) that of the main simulation box, with a simulation box size of $L=2048 \,\rm{Mpc/h}$. This case involves distinguishing between clustering DE with $w_{\rm DE}=-0.95$ and the cosmological constant scenario ($w_{\rm{DE}}=-1$) . 
  \label{b2}}
\end{figure}

In \Cref{fig6}, we investigate the accuracy of our ML algorithms while varying the number of sub-boxes used in the training dataset. We consider sub-boxes that are one-fourth (1/4) of the size of the main simulation box ($L=2048 \, \rm{Mpc/h}$) for distinguishing clustering DE with $w_{\rm DE}=-0.95$ from the cosmological constant scenario ($w_{\rm{DE}}=-1$). The results in \Cref{fig6} demonstrate that increasing the number of sub-boxes beyond 2000 does not significantly improve the algorithm's accuracy and justifies the number of sub-boxes chosen throughout this paper.

\section{Convolution Neural Network Architecture }\label{appendix1}

The CNN algorithm offers great flexibility in terms of its architecture, allowing us to adjust the number of hidden layers, neurons, and filters to suit our needs. In our study, we have designed a CNN architecture for the classification of cosmological models, specifically the clustering DE from the cosmological constant scenario. This architecture enables us to accurately categorise and analyse these cosmological scenarios.
To ensure optimal performance without overfitting, we adopt a progressive approach in our analyses. We start with a simple model and gradually increase its complexity, allowing it to capture the data more accurately.

In this study, we propose a CNN architecture tailored for processing volumetric data with an input shape of $64\times64\times64$ or $128\times128\times128$. 
The model comprises seven layers designed to systematically extract and capture features from the volumetric input. The initial layer is a 3D convolutional layer with a $7\times7\times7$ kernel and a stride of 2, which aims to capture local patterns in the input space while downsampling the spatial dimensions. Batch normalization is applied after each convolutional layer to improve training stability and convergence. Subsequent convolutional layers follow a similar pattern, gradually increasing the depth of feature maps to facilitate more abstract representations. Additionally, a max-pooling layer with a pooling size of $2\times2\times2$ is considered after forth convolutional layer to further downsample the spatial dimensions and enhance translation invariance. The model is regularised using the \textit{L2 regularization} \citep{a49} with a penalty coefficient of 0.01 applied to both the kernel weights and biases of the convolutional layers to prevent overfitting. Following the convolutional layers, a fully connected layer with 128 units with rectified linear units (ReLU) activation function \citep{agarap2019deep} is employed to capture high-level semantic information extracted from the feature maps. After that, a \textit{Dropout} \citep{a48} with a dropout rate of 0.5 is applied to mitigate overfitting in the fully connected layer. Finally, another fully connected layer with \textit{Sigmoid} activation function is employed in the output layer to produce binary classification predictions. For a detailed description of the architecture for the input shape of $64\times64\times64$ , refer to \autoref{b1}.

\begin{table}
    \centering
    \caption{CNN Model Architecture}
    \begin{tabular}{|c|c|c|c|}
        \hline
        \textbf{Layer Type} & \textbf{Output Shape} & \textbf{Parameters} \\
        \hline
        \hline
        Input & (64, 64, 64, 1) & 0 \\
        Conv3D (7x7x7), stride=2 & (32, 32, 32, 8) & 2752 \\
        BatchNormalization & (32, 32, 32, 8) & 32 \\
        Conv3D (7x7x7), stride=2 & (16, 16, 16, 16) & 43920 \\
        BatchNormalization & (16, 16, 16, 16) & 64 \\
        Conv3D (7x7x7), stride=2 & (8, 8, 8, 32) & 175648 \\
        BatchNormalization & (8, 8, 8, 32) & 128 \\
        Conv3D (7x7x7), stride=2 & (4, 4, 4, 16) & 175632 \\
        MaxPooling3D (2x2x2) & (2, 2, 2, 16) & 0 \\
        BatchNormalization & (2, 2, 2, 16) & 64 \\
        Conv3D (7x7x7), stride=2 & (1, 1, 1, 8) & 43912 \\
        BatchNormalization & (1, 1, 1, 8) & 32 \\
        Flatten & (8) & 0 \\
        Dense (128) & (128) & 1152 \\
        Dropout (0.5) & (128) & 0 \\
        Dense (1) & (1) & 129 \\
        \hline
        \multicolumn{2}{|c|}{\textbf{Total Parameters}} & 443465 \\ \hline
        \multicolumn{2}{|c|}{\textbf{Trainable Parameters}} & 443305 \\ \hline
        \multicolumn{2}{|c|}{\textbf{Non-Trainable Parameters}} & 160 \\
        \hline
    \end{tabular}
    \label{b1}
\end{table}

There are various performance measures that can be used to assess the training progress of our ML algorithms. One such measure is the loss function, which quantifies the difference between the algorithm's current output and the expected output. For our CNN algorithm, we utilise the \textit{binary cross-entropy loss} function, which is commonly used in training binary classifiers \citep{binary}.
\\
In the CNN architecture, our goal is to minimise the binary cross-entropy loss. To achieve this, we employ the \textit{Adam} optimiser \citep{kingma2017adam} with a learning rate of 0.0002 and a beta parameter ($\beta_1$) of 0.9. Additionally, we implement two checkpoints in the model. The first is the \textit{ModelCheckpoint}, which saves the best model during training based on a specified metric. The second is the \textit{ReduceLROnPlateau}, which reduces the learning rate when there is no progress in training. Moreover, we incorporate the \textit{EarlyStopping} callback, which stops the training process if the validation accuracy does not improve.
\\
To monitor the training progress, we record the loss function and accuracy computed on the training data after each epoch. We also calculate the validation loss and accuracy on a separate validation dataset. These metrics provide insights into the performance of the model during training and help us track its progress.

\section{Simple Classifier}\label{SC}
In this appendix, we present a simple classifier method based on a more traditional approach utilising the density field variance distribution of sub-boxes to differentiate between clustering DE and $\Lambda$CDM scenarios. The simplicity of these types of classifiers offers a more straightforward interpretation of the results compared to ML techniques. Additionally, it is notably easier to construct, serving as a valuable baseline to assess whether ML techniques outperform classical methods. 
\\
We analyse six simulation boxes, each with a fixed value of $N_{\rm grid} = 256^3$ and a box size of $L=2048 , \rm{Mpc/h}$ for every cosmology considered. Specifically, we consider values of $w$ ranging from $-0.7$ to $-0.99$ with intervals of $0.05$, while keeping $c_s^2$ fixed at $1$. To smooth the simulation boxes, we employ the \textit{Pylians3} library with a smoothing scale set to $8 \, \rm{Mpc/h}$.
Subsequently, we extract 2000 sub-boxes from each of the five density snapshots, resulting in 10,000 sub-boxes for each cosmology under consideration. Following this, we compute the variance of density field for each sub-box, generating a distribution of variances for each cosmology. Furthermore, we extract 2000 sub-boxes from the remaining snapshot, which we designate as our test dataset.
 
Afterwards, we assess the variance of each test sub-box, and measure the deviation of individual sub-box variances from the mean of the variance distribution linked to each cosmology.  If the variance of a sub-box is closer to the mean of the variance distribution for a particular model, we assign that model to the corresponding test sub-box. Thus, we establish a confusion matrix and evaluate the accuracy of this simple classifier method in comparison to our machine learning models.

\begin{figure}
\includegraphics[width=\columnwidth]{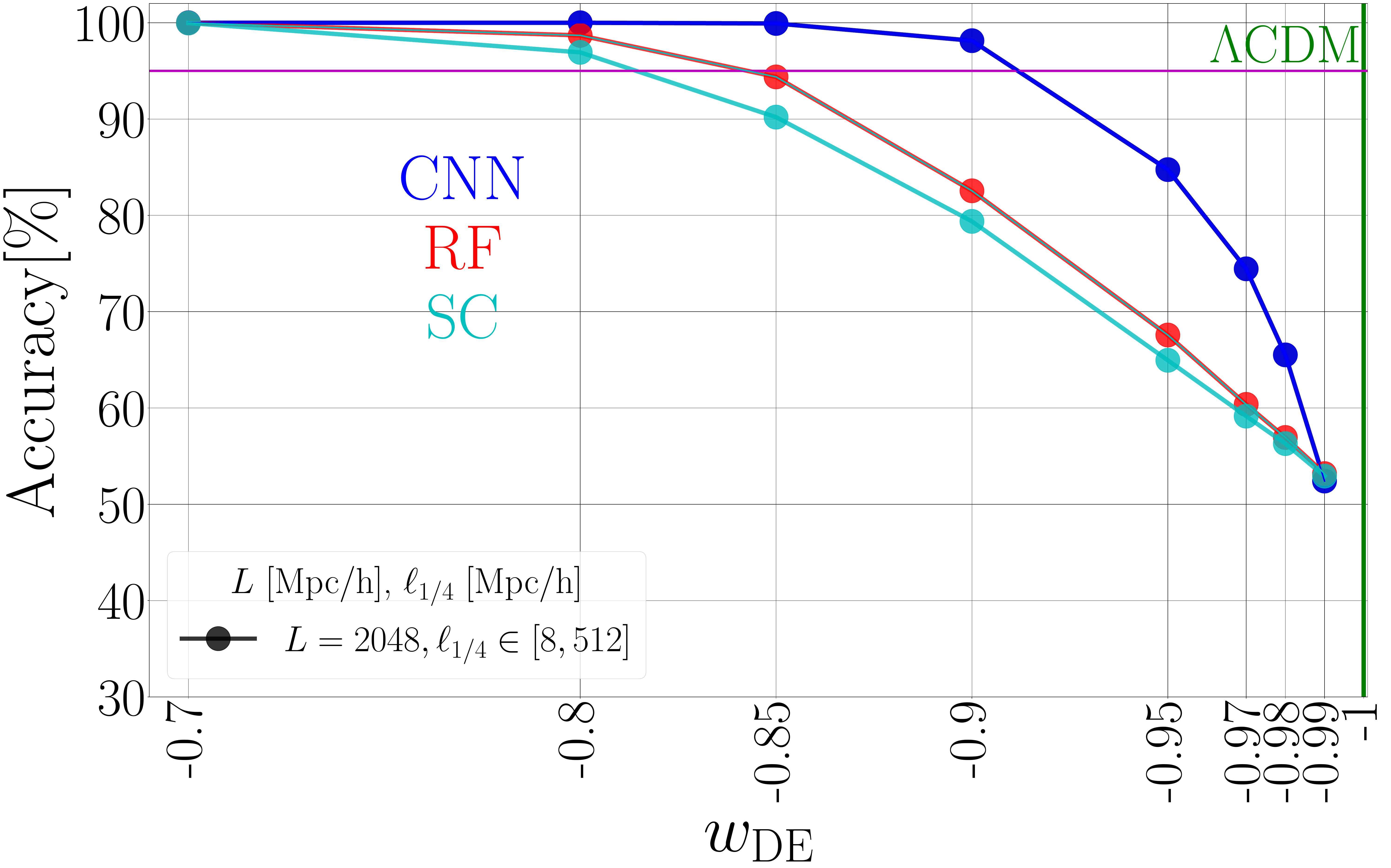}
\caption{ \label{fig7}
  The accuracy obtained from the CNN, RF and simple classifier (shown as SC) models for various values of $w_{\rm{DE}}$ is presented. The test sub-boxes have a size one-fourth (1/4) that of the main simulation box, with a simulation box size of $L=2048 \,\rm{Mpc/h}$.}
\end{figure}

In \Cref{fig7} we compare the accuracy of the CNN, RF, and simple classifier (SC) models for different values of $w_{\rm{DE}}$ and smoothing scale R equal to $8 \, \rm{Mpc/h}$. Our results show that the accuracy of the simple classifier is lower compared to the performance of RF and CNN models across all equations of state.

\end{document}